(* TautoPiEllPivf (the final version): no additional changes, only three additional references added [43,44,49]; published in JHEP02(2024)215 *)

\documentclass[aps,prd,nofootinbib,showpacs,preprintnumbers,amsmath,amssymb]{revtex4-1}
\usepackage{epsfig}
\usepackage{graphics}
\usepackage{graphicx}
\usepackage{amssymb}
\usepackage{latexsym}
\usepackage{multirow}
\usepackage{graphicx,color}
\def\be{\begin{equation}}
\def\ee{\end{equation}}
\def\bea{\begin{eqnarray}}
\def\eea{\end{eqnarray}}
\def\bes{\begin{subequations}}
\def\ees{\end{subequations}}
\def\gsim{~\rlap{$>$}{\lower 1.0ex\hbox{$\sim$}}\;}

\begin{document}

\title{Rare tau decays via exchange of on-shell almost degenerate Majorana neutrinos, $\tau^{\mp} \to \pi^{\mp} N_j \to \pi^{\mp} \mu^{\mp} \pi^{\pm}$ and  $\tau^{\mp} \to \pi^{\mp} N_j \to \pi^{\mp} \mu^{\pm} \pi^{\mp}$}

\author{Gorazd Cveti\v{c}$^a$}
 \email{gorazd.cvetic@usm.cl}
\author{ C.~S.~Kim$^b$}
\email{ cskim@yonsei.ac.kr}

\affiliation{$^a$ Department of Physics, Universidad T{\'e}cnica Federico Santa Mar{\'\i}a,  Casilla 110-V, Valpara{\'\i}so, Chile\\
  $^b$  Department of Physics and IPAP, Yonsei University, Seoul 120-749, Korea
  }

\begin{abstract}
  We investigate rare decays of tau leptons that occur via exchange of heavy on-shell neutrinos $N_j$ ($j=1,2$). These neutrinos can be either Dirac or Majorana, and are considered to be almost degenerate in mass. The decays can thus be either lepton number conserving (LNC), $\tau^{\mp} \to \pi^{\mp} N_j \to \pi^{\mp} \mu^{\mp} \pi^{\pm}$, or lepton number violating (LNV), $\tau^{\mp} \to \pi^{\mp} N_j \to \pi^{\mp} \mu^{\pm} \pi^{\mp}$. If neutrinos are Dirac, only LNC decays are possible. If they are Majorana, both LNC and LNV are possible. We derive the corresponding expressions for the effective decay widths $\Gamma_{{\rm eff},\mp}^{\rm (X)}$ (X=LNC, LNV) of these rare decays, where we account for $N_1$-$N_2$ overlap and oscillation effects and for the finite detector length effects. We then numerically evaluate these decay widths as well as the related CP violation asymmetry width $\Delta \Gamma_{\rm CP}^{\rm (X)} = (\Gamma_{{\rm eff},-}^{\rm (X)} - \Gamma_{{\rm eff},+}^{\rm (X)})$. We conclude that for certain, presently allowed, ranges of the heavy-light neutrino mixing parameters, such decays and asymmetries could be observed in Belle II experiment.
\end{abstract}
\maketitle

\section{Introduction}
\label{sec:Intro}

The success of the Standard Model (SM) of particle physics is not complete, as there exist theoretical problems (many parameters, finetuning) as well as experimental evidence which is not incorporated into it, such as Neutrino Oscillations \cite{Fukuda:1998mi,Eguchi:2002dm} and the related small nonzero masses of (light) neutrinos ($m_{\nu} < 1$ eV).

The models with See-Saw mechanism \cite{Minkowski:1977sc,Yanagida:1979as,Gell-Mann:1979vob} can explain the nonzero masses of three light neutrinos, and implies the existence of heavy (almost sterile) neutrinos as well as light neutrinos, and all these neutrinos are Majorana. The heavy neutrinos $N_j$ have strongly suppressed mixing with the SM-like flavour neutrinos $\nu_{\ell}$ ($\ell = e, \mu, \tau$). Majorana neutrinos can induce the lepton number conserving (LNC) and lepton number violating (LNV) processes, while Dirac neutrinos can induce only LNC processes. Heavy neutrinos can be searched in rare meson decays ~\cite{CPVBelle,Dib:2000wm,Cvetic:2012hd,GCCSKJZS1,GCCSKJZS2,symm,
Cvetic:2021itw,oscGCetal,Moreno:2016cfz,Milanes:2018aku,Mejia-Guisao:2017gqp,
Beltran:2022ast,Cvetic:2010rw,Dib:2019ztn,Dib:2018iyr,Dib:2016wge,Dib:2015oka}, at colliders~\cite{Milanes:2016rzr,Tapia:2021gne,Das:2018usr,Das:2017nvm,Das:2012ze,
Antusch:2017ebe,Das:2017rsu,Das:2017zjc,Chakraborty:2018khw,Cvetic:2019shl,
Antusch:2016ejd,Cottin:2018nms,Duarte:2018kiv,Drewes:2019fou,Cvetic:2018elt,
Cvetic:2019rms,Das:2018hph,Das:2016hof,Das:2017hmg,Antusch:2023jsa,Fernandez-Martinez:2022gsu,Najafi:2020dkp}, and in tau factories \cite{Zamora-Saa:2016ito,Kim:2017pra,Tapia:2019coy,Dib:2019tuj,Cheung:2020buy}.

A prominent neutrino mass model is the Neutrino-Minimal-Standard-Model ($\nu$MSM)~\cite{Asaka:2005an,Asaka:2005pn}. It is based on a variant of See-Saw mechanism and has two almost degenerate heavy neutrinos, with masses $M_{N1} \approx M_{N2} \sim 1 $ GeV, that can oscillate among themselves. 
Another specific model that has two almost degenerate heavy neutrinos with oscillation among themselves is considered in Ref.~\cite{Antusch:2023jsa}.
Those models can lead to a baryon asymmetry via leptogenesis \cite{Akhmedov:1998qx} and provides a natural Dark Matter candidate by a third heavy neutrino with mass $M_{N3} \sim 1$ keV. In addition, CP invariance must be broken in order to produce baryon asymmetry \cite{Sakharov:1967dj}.

Recent neutrino oscillation experiments indicate that mixing-angle $\theta_{13}$ is nonzero~\cite{An:2012eh}, and this opens the possibility of CP violation in the light neutrino sector \cite{T2K:2018rhz,RENO:2012mkc}. However, this CP-violation source is insufficient, and other sources of CP violation are required in order to explain the baryon asymmetry via leptogenesis (\cite{Chun:2017spz} and references therein). One such source is CP violation in the sector of heavy neutrinos, with masses $M_N < 246$ GeV~\cite{Akhmedov:1998qx,Drewes:2016gmt}.

In a previous work \cite{Cvetic:2021itw}, we considered such a scenario in the rare Higgs decays, $H \to \nu_k N_j \to \nu_k \ell {\bar q} q'$, where $N_j$ ($j=1,2$) are two almost degenerate in mass heavy on-shell neutrinos which can oscillate between themselves. In the present work, we consider such a scenario in the rare $\tau$ decays, $\tau^{\mp} \to \pi^{\mp} N_j \to \pi^{\mp} \mu \pi$, especially in view of the fact that many $\tau$'s will be produced in various experiments in the near future. A similar study was performed in Refs.~\cite{Zamora-Saa:2016ito,Tapia:2019coy}. We extend the latter studies in the sense that we now consider the $N_1$-$N_2$ overlap effects and the related $N_1$-$N_2$ neutrino oscillation effects in the overlap terms (similarly as in \cite{Cvetic:2021itw}). The suppression effects due to the finite length of the detector are now accounted for by explicit terms in the formulas. Further, the expressions are derived and numerical evaluations are performed for both the Majorana and the Dirac cases. On the other hand, another difference with the work \cite{Tapia:2019coy} is that the authors of  \cite{Tapia:2019coy}  took into account that $\tau$ leptons at Belle II experiment have a distribution of momenta in the lab frame, which they obtained by numerical simulation. In the present work, we did not use numerical simulations, instead used the averaged value of $\tau$ lepton momenta (i.e., the velocity $\beta_{\tau} = 0.9419$, cf.~Sec.~\ref{sec:res}).

In Sec.~\ref{sec:Gnoosc}, we study the mentioned rare decays of $\tau$ without oscillation effects. In Sec.~\ref{sec:Gosc} the oscillation effects are now included. In Sec.~\ref{sec:Geff} we include the effects of finite detector length in the decay width, i.e., we exclude from consideration the decays where the heavy on-shell neutrino does not decay within the detector. In Sec.~\ref{sec:res}, we present the results of numerical evaluations, for both the case of Majorana and of Direc neutrinos, and in Sec.~\ref{sec:concl} we summarise the results. In Appendices \ref{app:T2X01} and  \ref{app:d3} we present some explicit expressions for the considered decay amplitudes and for the final state integrations, respectively.

\section{Formula for  $\Gamma_{\mp} = \Gamma(\tau^{\mp} \to \pi^{\pm} N_j \to \pi^{\pm} \mu \pi)$, LNC and LNV, no oscillation effects}
\label{sec:Gnoosc}

We consider a scenario where we have, in addition to the three known light mass eigenstate neutrinos $\nu_k$ ($k=1,2,3$), two or more heavy neutrinos $N_j$ with masses $M_j \sim 10^1$ GeV. Further, we assume that two of the latter are almost mass degenerate, $M_1 \approx M_2$, in the mass regime $0.2 \ {\rm GeV} < M_j < 1.6 \ {\rm GeV}$ ($j=1,2$), thus kinematically allowing production of on-shell $N_j$'s in the above rare processes. The three flavour eigenstate neutrinos $\nu_{\ell}$ ($\ell=e, \mu, \tau$) are primarily a superposition of the three light neutrinos $\nu_k$, but with small admixtures $ U_{\ell N_j}$ of the heavy $N_j$'s
\be
\nu_{\ell} = \sum_{k=1}^3 U_{\ell \nu_k} \nu_k + \sum_{j=1}^2 U_{\ell N_j} N_j.
\label{mix} \ee
The upper bounds on the (squared) absolute values $|U_{\ell N_j}|^2$ of the heavy-light mixing coefficients $U_{\ell N_j}$ are determined principally by the present nondetection of various processes involving such heavy neutrinos (for reviews, see e.g.~\cite{Atre,Cvetic:2018elt}). The upper bounds in the mentioned mass range are very low (i.e., strong) for $|U_{e N_j}|^2$, i.e., when the flavour eigenstate is electron neutrino ($\nu_e$), primarily because of the nondetection of the neutrinoless double beta decay. Therefore, we will take in the rare $\tau$-decays $\tau \to \pi \ell \pi$ the produced charged lepton $\ell$ to be $\ell =\mu$.

Beside the known effective $W^{\ast} \pi$ couplings, the relevant couplings for the considered rare decays are then the $\ell N_j W$ couplings (where $\ell = \tau, \mu$):
\bes
\label{ellNW}
\bea
{\cal L}_{\ell N_j W} & = & - \left( \frac{g}{2 \sqrt{2}} \right) \sum_{j=1}^2
\left\{
U _{\ell N_j} {\overline {\ell}} \gamma^{\eta} (1 - \gamma_5) N_j W^{-}_{\eta} +
U_{\ell N_j}^{\ast} {\overline {N_j}}  \gamma^{\eta} (1 - \gamma_5) \ell W^{+}_{\eta}
\right\}
\label{ellNWa} \\
& = &  + \left( \frac{g}{2 \sqrt{2}} \right) \sum_{j=1}^2
\left\{
U _{\ell N_j} {\overline {N_j^c}} \gamma^{\eta} (1 + \gamma_5) \ell^c W^{-}_{\eta} +
U_{\ell N_j}^{\ast} {\overline {\ell^c}}  \gamma^{\eta} (1 + \gamma_5) N_j^c W^{+}_{\eta}
\right\}
\label{ellNWb}
\eea \ees
Here, the charged-conjugated field is $N^c$ (equal to $- i \gamma^2 N^{\ast}$ in the Dirac and chiral representations). When $N_j$'s are Majorana, we have $N_j^c=N_j$.

\begin{figure}[htb] 
\centering\includegraphics[width=100mm]{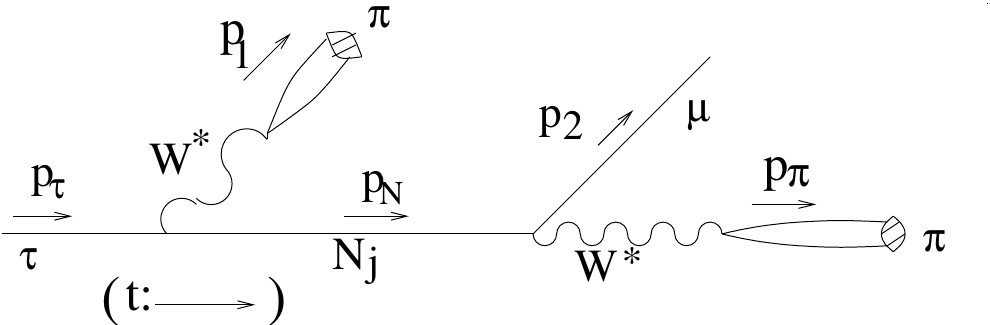}
\caption{\footnotesize The decay $\tau^{\mp}(p_{\tau}) \to \pi^{\mp}(p_1) N_j(p_N) \to \pi^{\mp}(p_1) \mu(p_2) \pi(p_{\pi})$, where $N_j$ is considered on-shell.}
\label{Figtaudec}
\end{figure}
Direct evaluation then gives, for the LNC processes $\tau^{\mp} \to \pi^{\mp} N_j \to \pi^{\mp} \mu^{\mp} \pi^{\pm}$, cf.~Fig.\ref{Figtaudec}, the following respective ${\cal T}_{\mp}^{\rm (LNC)}$ reduced scattering amplitudes:
\bes
\label{TLC}
\bea
{\cal T}_{-}^{\rm (LNC)} & = & - \frac{1}{2} K^2 \sum_{j=1}^2 U_{\tau N_j}^{\ast} U_{\mu N_j} \left[ {\bar u}_{(\mu)}(p_2,h_2) {\displaystyle{\not}p}_{\pi} (1 - \gamma_5) ({\displaystyle{\not}p}_{\tau} - {\displaystyle{\not}p}_1+M_{N_j}) {\displaystyle{\not}p}_1 (1 - \gamma_5) u_{(\tau)}(p_{\tau},h_{\tau}) \right] P_j(p_N^2),
\label{TLCm} \\
{\cal T}_{+}^{\rm (LNC)} & = & + \frac{1}{2} K^2 \sum_{j=1}^2 U_{\tau N_j} U_{\mu N_j}^{\ast} \left[ {\bar v}_{(\tau)}(p_{\tau},h_{\tau}) {\displaystyle{\not}p}_1 (1 - \gamma_5) (- {\displaystyle{\not}p}_{\tau} + {\displaystyle{\not}p}_1+M_{N_j}) {\displaystyle{\not}p}_{\pi} (1 - \gamma_5) v_{(\mu)}(p_2,h_2) \right] P_j(p_N^2).
\label{TLCp}
\eea \ees
Here,
\be
K \equiv G_F f_{\pi} V_{u d},
\label{Kdef} \ee
where $G_F=1.1664 \times 10^{-5} \ {\rm GeV}^{-2}$ is the Fermi coupling constant, $f_{\pi}=0.1304$ GeV is the pion decay constant. Further, we denoted by $P_j(p_N^2)$ the denominator of the $N_j$ propagator
\be
P_j(p_N^2) \equiv \frac{1}{( p_N^2 - M_{N_j}^2 + i \Gamma_{N_j} M_{N_j})} \equiv P_j.
\label{Pj} \ee
and $p_N^2 = (p_{\tau}-p_1)^2$. The symbols $h_2$ and $h_{\tau}$ in Eqs.~(\ref{TLC}) denote the helicities of muon and tau.

Analogously, the ${\cal T}_{\mp}^{\rm (LNV)}$ scattering amplitudes for the LNV processes $\tau^{\mp} \to \pi^{\mp} N_j \to \pi^{\mp} \mu^{\pm} \pi^{\mp}$ are
\bes
\label{TLV}
\bea
{\cal T}_{-}^{\rm (LNV)} & = & + \frac{1}{2} K^2 \sum_{j=1}^2 U_{\tau N_j}^{\ast} U_{\mu N_j}^{\ast} \left[ {\bar v}_{(\tau)}(p_{\tau},h_{\tau}) {\displaystyle{\not}p}_1 (1 + \gamma_5) (- {\displaystyle{\not}p}_{\tau} + {\displaystyle{\not}p}_1+M_{N_j}) {\displaystyle{\not}p}_{\pi} (1 - \gamma_5) v_{(\mu)}(p_2,h_2) \right] P_j(p_N^2),
\label{TVLm} \\
{\cal T}_{+}^{\rm (LNV)} & = & + \frac{1}{2} K^2 \sum_{j=1}^2 U_{\tau N_j} U_{\mu N_j}
\left[ {\bar u}_{(\mu)}(p_2,h_2) {\displaystyle{\not}p}_{\pi} (1 - \gamma_5) ({\displaystyle{\not}p}_{\tau} - {\displaystyle{\not}p}_1+M_{N_j}) {\displaystyle{\not}p}_1 (1 + \gamma_5) u_{(\tau)}(p_{\tau},h_{\tau}) \right] P_j(p_N^2).
\label{TLVp}
\eea \ees
We notice from Eqs.~(\ref{TLC}) and (\ref{TLV}) that, due to the chirality factors $(1 \pm \gamma_5)$, in the numerator factor $[ \pm ( {\displaystyle{\not}p}_{\tau} - {\displaystyle{\not}p}_1) + M_{N_j}]$ of the $N_j$-propagator the term $M_{N_j}$ does not contribute in the LNC case, and the term $( {\displaystyle{\not}p}_{\tau} - {\displaystyle{\not}p}_1)$ does not contribute in the LNV case.

Before squaring the above amplitudes (and summing over helicities $h_2$ and averaging over helicities $h_{\tau}$), we take into account that the two intermediate hevay neutrinos $N_j$ in our scenario are nearly degenerate in mass
\bes
\label{massdeg}
\bea
M_{N_1} &\equiv & M_N; \quad \Delta M_N \equiv M_{N_2} - M_{N_1};
\label{MNdef} \\
&& 0 < \Delta M_N (\lesssim \Gamma_{N_j}) \ll M_N.
\label{deg} \eea \ees
As a consequence, the quadratic terms in the intermediate neutrino propagators $P_j(p_N^2) \equiv P_j$ can be written as
\bes
\label{PP}
\bea
P_j P_j^{\ast} &=& \frac{\pi}{M_N \Gamma_{N_j}} \delta(p_N^2 - M_N^2) \quad (j=1,2),
\label{PjPj} \\
{\rm Im} (P_1 P_2^{\ast}) & = & \frac{\eta(y)}{y} \frac{\pi}{M_N \Gamma_{N}} \delta(p_N^2 - M_N^2),
\label{ImP1P2} \\
{\rm Re} (P_1 P_2^{\ast}) & = & \delta(y) \frac{\pi}{M_N \Gamma_{N}} \delta(p_N^2 - M_N^2),
\label{ReP1P2}
\eea \ees
where
\bes
\label{yetdel}
\bea
y & = &\frac{\Delta M_N}{\Gamma_N},  \qquad \Gamma_N = \frac{1}{2} ( \Gamma_{N_1} + \Gamma_{N_2} ),
\label{yGN} \\
\frac{\eta(y)}{y} & = & \frac{y}{1+y^2},
\label{eta} \\
\delta(y) &=& \frac{1}{1 + y^2}.
\label{del} \eea \ees
The expressions Eqs.~(\ref{eta})-(\ref{del}) were obtained first numerically in rare decay processes of pseudoscalar mesons via almost degenerate neutrinos $N_j$ ($j=1,2$) in Refs.~\cite{GCCSKJZS1,GCCSKJZS2}. A derivation of the expression (\ref{eta}) was presented in Ref.~\cite{symm} (App.~6 there), and of the expression (\ref{del}) in Ref.~\cite{Cvetic:2021itw} (App.~B there).

This implies, among other things, that the squaring of the amplitudes (\ref{TLC}) and (\ref{TLV}) is to be performed now by taking into account that $p_N^2 = (p_{\tau}-p_{1})^2$ is replaced by $M_N^2$  (i.e., $p_N^2 \mapsto M_N^2$). This then implies
\bes
\label{sqT}
\bea
\langle | {\cal T}_{-}^{\rm (LNC)}|^2 \rangle & = &  \frac{1}{2} K^4
\sum_{i=1}^2 \sum_{j=1}^2 U_{\tau N_j}^{\ast} U_{\tau N_i} U_{\mu N_j} U_{\mu N_i}^{\ast} P_j P_i^{\ast} T_2^{(\rm LNC)},
\label{sqTLCm} \\
\langle | {\cal T}_{+}^{\rm (LNC)}|^2 \rangle & = &  \frac{1}{2} K^4
\sum_{i=1}^2 \sum_{j=1}^2 U_{\tau N_j} U_{\tau N_i}^{\ast} U_{\mu N_j}^{\ast} U_{\mu N_i} P_j P_i^{\ast} T_2^{(\rm LNC)},
\label{sqTLCp} \\
\langle | {\cal T}_{-}^{\rm (LNV)}|^2 \rangle & = &  \frac{1}{2} K^4
\sum_{i=1}^2 \sum_{j=1}^2 U_{\tau N_j}^{\ast} U_{\tau N_i} U_{\mu N_j}^{\ast} U_{\mu N_i} P_j P_i^{\ast} T_2^{(\rm LNV)},
\label{sqTLVm} \\
\langle | {\cal T}_{+}^{\rm (LNV)}|^2 \rangle^{\ast} & = &  \frac{1}{2} K^4
\sum_{i=1}^2 \sum_{j=1}^2 U_{\tau N_j} U_{\tau N_i}^{\ast} U_{\mu N_j} U_{\mu N_i}^{\ast} P_j P_i^{\ast} T_2^{(\rm LNV)},
\label{sqTLVp}
\eea \ees
where $\langle \ldots \rangle$ denotes sum over $h_2$ helicities of muon and average over the $h_{\tau}$ helicities of tau, and the expressions $T_2^{\rm (X)}$ (X=LNC, LNV) are
\bes
\label{T2}
\bea
T_2^{(\rm LNC)} & = & 2 {\rm Tr} \left[{\displaystyle{\not}p}_{\tau} {\displaystyle{\not}p}_1 {\displaystyle{\not}p}_N {\displaystyle{\not}p}_{\pi} {\displaystyle{\not}p}_2 {\displaystyle{\not}p}_{\pi}  {\displaystyle{\not}p}_N {\displaystyle{\not}p}_1 (1 - \gamma_5) \right] = T_2^{(\rm LNC,0)} + (p_1 \cdot p_2) T_2^{(\rm LNC,1)},
\label{T2LC} \\
T_2^{(\rm LNV)} & = & 2 M_N^2 {\rm Tr} \left[{\displaystyle{\not}p}_{\tau} {\displaystyle{\not}p}_1 {\displaystyle{\not}p}_{\pi} {\displaystyle{\not}p}_2 {\displaystyle{\not}p}_{\pi}  {\displaystyle{\not}p}_1 (1 +\gamma_5) \right] = T_2^{(\rm LNV,0)} + (p_1 \cdot p_2) T_2^{(\rm LNV,1)},
\label{T2LV}
\eea \ees
where $p_N^2 = (p_{\tau}-p_1)^2 \mapsto M_N^2$, and the explicit expressions $T_2^{(\rm X,0)}$ and $T_2^{(\rm X,1)}$ are given in Appendix \ref{app:T2X01}.

The decay widths of the considered decays are obtained by integrating the above squares over the final phase space
\bea
d \Gamma( \tau \to \pi \mu \pi) & = & \frac{1}{ 2 M_{\tau} (2 \pi)^5} |{\cal T}_{\mp}^{\rm (X)}|^2 \; d_3,
\label{dGamma1} \eea
where $d_3$ is the differential of the integration over the phase space of the (three) final particles
\be
\label{d3}
d_3  =   d_2(\tau \to \pi N_j) \; d p_N^2 \; d_2(N_j \to \mu \pi).
\ee
We note that this integration can be performed analytically, giving for the terms in Eqs.~(\ref{T2}) at unity (at $T_2^{\rm (X,0)}$) one specific expression, and for the terms at $(p_1 \cdot p_2)$ (at $T_2^{\rm (X,1)}$) another specific expression, cf.~ Appendix \ref{app:d3}. After performing these integrations, a factorisation of the obtained expression for the decay width can be performed, resulting in
\bes
\label{Gamnoosc}
\bea
\Gamma(\tau^{\mp} \to \pi^{\mp} N_j \to \pi^{\mp} \mu^{\mp} \pi^{\pm})^{\rm (LNC)} & = &
      {\overline \Gamma}(\tau \to \pi \mu \pi) {\bigg \{}
       \frac{\Gamma_N}{\Gamma_{N_1}} |U_{\tau N_1}|^2 |U_{\mu N_1}|^2 +
      \frac{\Gamma_N}{\Gamma_{N_2}} |U_{\tau N_2}|^2 |U_{\mu N_2}|^2 +
      \nonumber\\
     && + 2 |U_{\tau N_1} U_{\tau N_2} U_{\mu N_1} U_{\mu N_2}|
      \left[ \delta(y) \cos \theta_{21} \pm \frac{\eta(y)}{y} \sin \theta_{21} \right] {\bigg \}},
      \label{GamnooscLC} \\
\Gamma(\tau^{\mp} \to \pi^{\mp} N_j \to \pi^{\mp} \mu^{\pm} \pi^{\mp})^{\rm (LNV)} & = &
      {\overline \Gamma}(\tau \to \pi \mu \pi) {\bigg \{}
       \frac{\Gamma_N}{\Gamma_{N_1}} |U_{\tau N_1}|^2 |U_{\mu N_1}|^2 +
      \frac{\Gamma_N}{\Gamma_{N_2}} |U_{\tau N_2}|^2 |U_{\mu N_2}|^2 +
      \nonumber\\
   &&   + 2 |U_{\tau N_1} U_{\tau N_2} U_{\mu N_1} U_{\mu N_2}|
      \left[ \delta(y) \cos \chi_{2 1} \mp \frac{\eta(y)}{y} \sin \chi_{21} \right] {\bigg \}}.
      \label{GamnooscLV}
\eea \ees
Here, the angles $\theta_{21}$ and $\xi_{21}$ are related to the phases of the heavy-light mixing parameters
\bes
\label{phases}
\bea
\psi_{\mu, j} & \equiv & {\rm Arg}(U_{\mu N_j}), \quad
\psi_{\tau,j}  \equiv {\rm Arg}(U_{\tau N_j}),
\label{psi} \\
\theta_{21} & = & (\psi_{\mu, 2} - \psi_{\tau, 2}) - (\psi_{\mu, 1} - \psi_{\tau, 1}),
\label{theta} \\
\chi_{21} & = & (\psi_{\mu, 2} + \psi_{\tau, 2}) - (\psi_{\mu, 1} + \psi_{\tau, 1}).
\label{xi}
\eea \ees
Furthermore, ${\overline \Gamma}(\tau \to \pi \mu \pi)$ in Eqs.~(\ref{Gamnoosc}) is the canonical decay width\footnote{Canonical stands for the case when the heavy-light factors $U_{\tau N_j}$ and $U_{\mu N_j}$ are replaced by unity.} which can be written in factorised form
\bea
{\overline \Gamma}(\tau \to \pi \mu \pi) & = & \frac{1}{\Gamma_N} {\overline \Gamma}(\tau \to \pi  N) {\overline \Gamma}(N \to \mu \pi),
\label{bGpimupi} \eea
where
\bes
\label{bGs}
\bea
{\overline \Gamma}(\tau \to \pi  N) & = & \frac{1}{16 \pi} K^2 M_{\tau}^3
\lambda^{1/2} \left( 1, \frac{M_N^2}{M_{\tau}^2}, \frac{M_{\pi}^2}{M_{\tau}^2} \right) \left[ \left( 1 -  \frac{M_N^2}{M_{\tau}^2} \right)^2 - \frac{M_{\pi}^2}{M_{\tau}^2} \left( 1 +  \frac{M_N^2}{M_{\tau}^2} \right) \right],
\label{bGtaupiN}
\\
{\overline \Gamma}(N \to \mu \pi) & = & \frac{1}{16 \pi} K^2 M_N^3
\lambda^{1/2} \left( 1, \frac{M_{\mu}^2}{M_N^2}, \frac{M_{\pi}^2}{M_N^2} \right) \left[ \left( 1 -  \frac{M_{\mu}^2}{M_N^2} \right)^2 - \frac{M_{\pi}^2}{M_N^2} \left( 1 +  \frac{M_{\mu}^2}{M_N^2} \right) \right].
\label{bGNmupi}
\eea \ees
Furthermore, in Eqs.~(\ref{Gamnoosc}) the quantities $\Gamma_{N_1}$ and $\Gamma_{N_2}$ are the total decay widths of $N_1$ and $N_2$, and $\Gamma_N$ is the arithmetic average thereof, i.e., the quantities that appeared already in Eqs.~(\ref{PP})-(\ref{yetdel}).

\begin{figure}[htb] 
\begin{minipage}[b]{.49\linewidth}
\includegraphics[width=85mm,height=50mm]{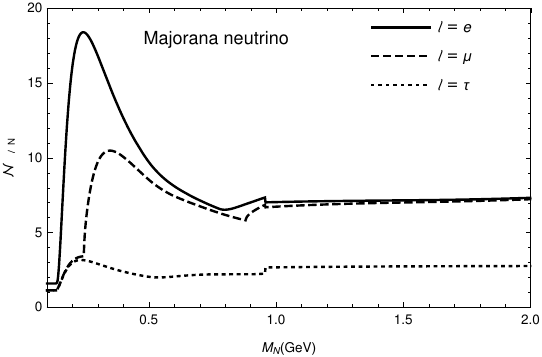}
\end{minipage}
\begin{minipage}[b]{.49\linewidth}
\includegraphics[width=75mm,height=47mm]{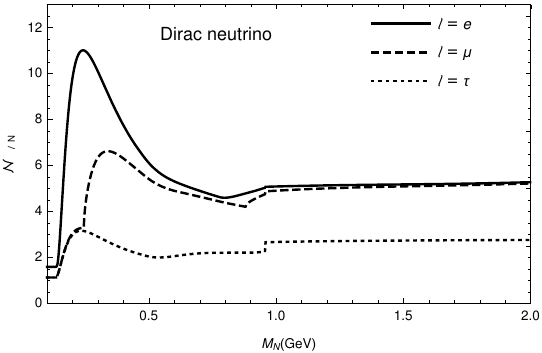}
\end{minipage} \vspace{12pt}
\caption{\footnotesize (a) The factors ${\cal N}_{\ell N}$ ($\ell = e, \mu, \tau$) appearing in Eq.~(\ref{calKj}), as a function of neutrino mass $N$, when $N$ is Majorana neutrino; (b) the same, but for the case when $N$ is Dirac neutrino.}
\label{FigGN}
\end{figure}

The total decay width $\Gamma_{N_j}$ of the neutrino contains the heavy-light mixing coefficients (for details, cf.~\cite{GCCSKJZS2,symm} and references therein)
\be
\Gamma_{N_j} = {\widetilde {\cal K}}_j(M_N) {\overline \Gamma}_N(M_N).
\label{GammaNj} \ee
Here,
\be
 {\overline \Gamma}_N(M_N) \equiv \frac{G_F^2 M_N^5}{96 \pi^3},
 \label{bGN} \ee
is the canonical factor. On the other hand, the factor ${\widetilde {\cal K}}_j$ ($j=1,2$) incorporates the heavy-light mixing coefficients
\be
 {\widetilde {\cal K}}_j(M_N) = {\cal N}_{e N} |U_{e N_j}|^2 +{\cal N}_{\mu N} |U_{\mu N_j}|^2 +{\cal N}_{\tau N} |U_{\tau N_j}|^2 \qquad (j=1,2).
\label{calKj} \ee
The factors ${\cal N}_{\ell N} = {\cal N}_{\ell N}(M_N) \sim 10^1$  ($\ell=e, \mu, \tau$) are effective mixing coefficients. They were evaluated in \cite{GCCSKJZS2,symm}. The factors ${\cal N}_{\ell N}(M_N)$, as a function of the $N_j$ neutrino mass $M_N$, are presented in Figs.~\ref{FigGN}(a), (b) for the case of the Majorana and Dirac neutrinos.
In these Figures we can see that in mass regime $M_N \approx 0.2$-$0.4$ GeV the factors ${\cal N}_{\ell N}$ for light leptons $\ell=e, \mu$ vary significantly with $M_N$. This has to do with the fact that in this mass regime there are important $N$ decay channels to the single light pseudoscalar and vector mesons, in addition to the pure lepton channels. In Figs.~\ref{FigGN} we see a small kink at $M_N=M_{\eta'}$ ($=0.9578$ GeV) because, for $M_N > M_{\eta'}$, due to duality the numerous semimesonic decay modes are evaluated as quark-antiquark decay modes, cf.~also \cite{GKS,HKS}. The fact that $\Gamma_N$ for Majorana neutrinos is larger than for Dirac neutrinos is due to the fact that the Majorana neutrinos have more decay channels; namely, for any existing decay channel of a neutrino, the charge-conjugated channel also exists if the neutrino is Majorana.

\section{Inclusion of oscillation effects}
\label{sec:Gosc}

At this point we include the oscillation effects between the two heavy quasidegenerate neutrinos $N_1$ and $N_2$. Based on the approach of Ref.~\cite{CGL}, it was argued in Ref.~\cite{oscGCetal} that the (quasidegenerate) $N_1$ and $N_2$ neutrinos, during their propagation between the two vertices of the decay process, have oscillation effects which are represented by the following replacements in the amplitudes of the considered decays $\tau \to \pi N \to \pi \mu \pi$:
\bes
\label{oscrepl}
\bea
U_{\mu N_j}       & \mapsto & U_{\mu N_j}        \exp(- i p_{N_j} \cdot z),
\label{Toscrepl1} \\
U_{\mu N_j}^{\ast} & \mapsto & U_{\mu N_j}^{\ast} \exp(- i p_{N_j} \cdot z) \qquad (j=1,2).
\label{Toscrepl2}
\eea \ees
The replacements are performed wherever these elements appear in the (reduced) matrix elements ${\cal T}$ of the corresponding decays. Hence, the first replacement is performed in ${\cal T}^{\rm (LNC)}(\tau^-)$ and ${\cal T}^{\rm (LNV)}(\tau^+)$, and the second replacement in the other two matrix elements, ${\cal T}^{\rm (LNC}(\tau^+)$ and ${\cal T}^{\rm (LNV)}(\tau^-)$. In all these replacements, $z=(t,0,0,L)$ is the 4-vector of distance between the two vertices; $p_{N_1}$ and $p_{N_2}$ are the momenta of the two neutrinos, and they differ slighly because of the (small) mass difference $\Delta M_N = M_{N_2}-M_{N_1}$. The difference in phases can be expressed as \cite{CGL,oscGCetal}
\be
(p_{N_2}-p_{N_1}) \cdot z = 2 \pi \frac{L}{L_{\rm osc}},
\label{difosc} \ee
where $L_{\rm osc}$ is the effective oscillation length
\be
L_{\rm osc} = \frac{2 \pi \beta_N \gamma_N}{\Delta M_N} \quad \Rightarrow \quad
\frac{2 \pi}{L_{\rm osc}} = y \frac{\Gamma_N}{\beta_N \gamma_N} .
\label{Losc} \ee
Here, $\beta_N = v_N/c$ where $v_N$ is the speed of the on-shell neutrinos $N_j$ in the lab (it is practically equal for $N_1$ and $N_2$) and $\gamma_N = 1/\sqrt{1 - \beta_N^2}$ is the corresponding Lorentz factor. We use the usual units where $c=1$ (and $\hbar=1$).

An approximate approach, using the replacements (\ref{oscrepl}) in amplitudes for some of the terms in squared amplitudes $\langle | {\cal T}|^2 \rangle$, was applied in Ref.~\cite{CPVBelle}. However, the use of these replacements in amplitudes for all terms in $\langle | {\cal T}|^2 \rangle$ [including the so called overlap terms $\propto \delta(y), \eta(y)$] was applied in Ref.~\cite{Cvetic:2021itw}. We apply this latter approach because it is consistent. It can then be verified that the use of these replacements (\ref{oscrepl}) and relations (\ref{difosc})-(\ref{Losc}) results in the relations (\ref{Gamnoosc}) having the replacements
\bes
\label{csrep}
\bea
\Gamma(\tau^{\mp})^{\rm (LNC)}: \;
\cos \theta_{21} & \mapsto & \cos\left( \frac{2 \pi L}{L_{\rm osc}} \mp \theta_{21} \right), \quad  \pm \sin \theta_{21} \mapsto - \sin \left( \frac{2 \pi L}{L_{\rm osc}} \mp \theta_{21} \right);
\label{csrepLC}
\\
\Gamma(\tau^{\mp})^{\rm (LNV)}: \;
\cos \chi_{21} & \mapsto & \cos\left( \frac{2 \pi L}{L_{\rm osc}} \pm \chi_{21} \right), \quad  \mp \sin \chi_{21} \mapsto - \sin \left( \frac{2 \pi L}{L_{\rm osc}} \pm \chi_{21} \right).
\label{csrepLV} \eea \ees
These results can be summarised in the following a more compact form:
\bea
\label{Gosc}
\Gamma (\tau^{\mp} \to \pi^{\mp} \mu \pi)^{\rm (X)} & = &
{\overline \Gamma}(\tau \to \pi \mu \pi)
 {\bigg \{}
       \sum_{k=1}^2 \frac{\Gamma_N}{\Gamma_{N_k}} |U_{\tau N_k}|^2 |U_{\mu N_k}|^2
      \nonumber\\
   &&   + 2 |U_{\tau N_1} U_{\tau N_2} U_{\mu N_1} U_{\mu N_2}|
      \left[ \delta(y) \cos  \left( \frac{2 \pi L}{L_{\rm osc}} \mp \Delta \Psi^{\rm (X)} \right) -  \frac{\eta(y)}{y} \sin \left( \frac{2 \pi L}{L_{\rm osc}} \mp \Delta \Psi^{\rm (X)} \right)  \right] {\bigg \}},
\eea
where X=LNC or X=LNV, and the corresponding phase differences are
\bes
\label{Delpsi}
\bea
\Delta \Psi^{\rm (LNC)} &=& (\psi_{\tau, 1}-\psi_{\mu, 1})-(\psi_{\tau, 2}-\psi_{\mu, 2})
\; (= \theta_{21}),
\label{delPsiLC} \\
\Delta \Psi^{\rm (LNV)} &=& (\psi_{\tau, 1}+\psi_{\mu, 1})-(\psi_{\tau, 2}+\psi_{\mu, 2})
\; (= - \chi_{21}),
\label{delPsiLV} \eea \ees
where $\psi_{\mu,j}$ and $\psi_{\tau,j}$ are the phases (arguments) of the heavy-light mixing elements $U_{\mu N_j}$ and $U_{\tau N_j}$, cf.~Eqs.~(\ref{phases}).

\section{Effective decay widths due to the detector finite length effects}
\label{sec:Geff}

We implicitly assumed that the two vertices of the considered decay $\tau \to \pi N_j \to \pi \mu \pi$ are always within the detector, or equivalently, that the detector is infinitely large. In practice, the detector has a finite length  $L_{\rm det}$ ($\sim 1$ m). Therefore, we need to exclude, in an effective approximate way, the decays where the on-shell neutrinos $N_j$ travel a distance $L > L_{\rm det}$ before decaying into $\mu \pi$. This would then give us a realistic, effective, decay width $\Gamma_{\rm eff}(L)$ as a function of the maximal length $L$ between the two vertices (where $L$ can be regarded as a variable). The resulting differential of the effective decay width, $d \Gamma_{\rm eff}(L) =  \Gamma_{\rm eff}(L+dL)-\Gamma_{\rm eff}(L)$, is then
\bea
\lefteqn{
  d \Gamma_{\rm eff}(\tau^{\mp} \to \pi^{\mp} \mu \pi; L)^{\rm (X)} =
{\overline \Gamma}(\tau \to \pi \mu \pi)  {\bigg \{}
\frac{\Gamma_N}{\Gamma_{N_1}} d P_{N_1}(L) |U_{\tau N_1}|^2 |U_{\mu N_1}|^2  +
\frac{\Gamma_N}{\Gamma_{N_2}} d P_{N_2}(L) |U_{\tau N_2}|^2 |U_{\mu N_2}|^2  +
}
\nonumber\\ &&
+
d P_{N}(L) 2  |U_{\tau N_1}| |U_{\tau N_2}| |U_{\mu N_1}| |U_{\mu N_2}|
\left[ \delta(y) \cos  \left( \frac{2 \pi L}{L_{\rm osc}} \mp \Delta \Psi^{\rm (X)} \right) -  \frac{\eta(y)}{y} \sin \left( \frac{2 \pi L}{L_{\rm osc}} \mp \Delta \Psi^{\rm (X)} \right)  \right] {\bigg \}}.
\label{dGeff1} \eea
Here, $dP_{N_j}(L) = P_{N_j}(L+d L) - P_{N_j}(L)$, where $P_{N_j}(L)$ is the probability that the propagating on-shell neutrino $N_j$ decays within the distance $L$ from its birth vertex
\bes
\label{PNjs}
\bea
P_{N_j}(L) &=& 1 - \exp \left( - \frac{L \Gamma_{N_j}}{\beta_N \gamma_N} \right),
\label{PNj} \\
d P_{N_j}(L) &=& \frac{\Gamma_{N_j}}{\beta_N \gamma_N} \exp \left( - \frac{L \Gamma_{N_j}}{\beta_N \gamma_N} \right) d L.
\label{dPNj} \eea \ees
We recall that $\Gamma_{N_j}$ is the total decay width of $N_j$ and $\gamma_N = (1 - \beta_N)^{-1/2}$ is the Lorentz lab time dilation factor. A priori, it is not clear which value of $\Gamma_{N_j}$ we should use in $d P_{N}(L)$ at the $N_1$-$N_2$ overlap contributions $\propto \delta(y), \eta(y)/y$ in Eq.~(\ref{dGeff1}). We propose to use at this point the average value $\Gamma_N$ (\ref{yGN}). In practical evaluations here, however, this will not matter, because we will assume from now on that $\Gamma_{N_1} = \Gamma_{N_2}$ (i.e., that $|U_{\ell_s N_1}|= |U_{\ell_s N_2}|$ for all $\ell_s$).

\begin{figure}[htb] 
\centering\includegraphics[width=100mm]{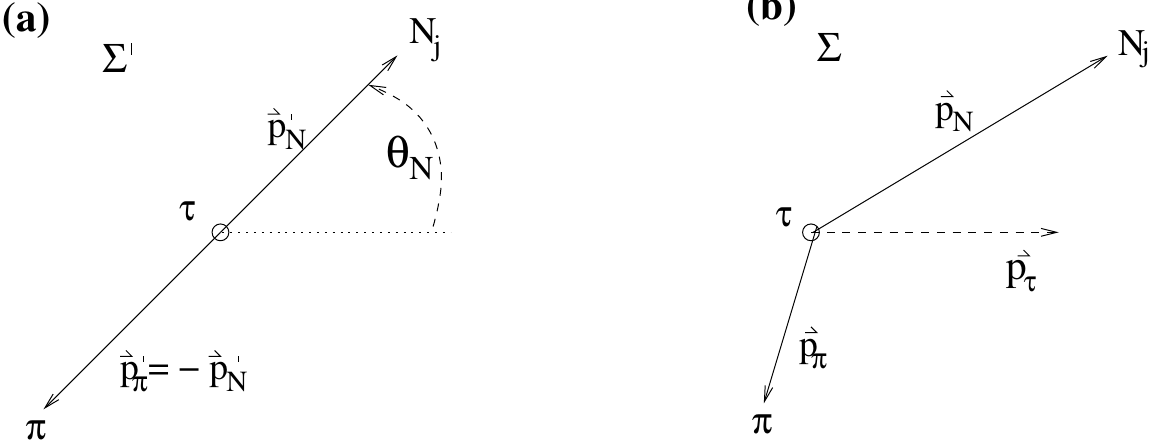}
\caption{\footnotesize (a) The 3-momenta of the produced particles in the decay $\tau \to N_j \pi$ in the $\tau$ rest frame ($\Sigma^{'}$); (b) the same momenta, but in the lab frame ($\Sigma$).}
\label{FigSpS}
 \end{figure}

Hence, from now on, we adopt the following simplifying assumptions:
\be
|U_{\ell N_1}| = |U_{\ell N_2}|  \; (= |U_{\ell N}|) \qquad (\ell = e, \mu, \tau),
\label{simpl} \ee
implying that the total decay widths of $N_1$ and $N_2$ are equal
\be
\Gamma_{N_1} = \Gamma_{N_2} = \Gamma_{N}; \qquad dP_{N_j}(L) =dP_N(L) \quad (j=1,2).
\label{dPN} \ee
In this case, the $L$-dependence of the expression (\ref{dGeff1}) simplifies. If, in addition, at first we assume that the lab speed $\beta_N$ of $N_j$ neutrinos is fixed, we can integrate the expression (\ref{dGeff1})\footnote{We first rename $L$ and $dL$ in the expression (\ref{dGeff1}) as $L'$ and $dL'$.}  over $d L'$ from $L'=0$ to $L'=L$, and obtain
\bes
\label{Geffb}
\bea
\lefteqn{
\Gamma_{\rm eff}(\tau^{\mp} \to \pi^{\mp} \mu \pi; L)^{\rm (X)} =
\int_{0}^{L} d L' \frac{d \Gamma_{\rm eff}(\tau^{\mp} \to \pi^{\mp} \mu \pi; L')}{d L'}
}
  \label{Geffbdef} \\
  & = &
{\overline \Gamma}(\tau \to \pi \mu \pi)   |U_{\tau N}|^2  |U_{\mu N}|^2 {\Bigg \{}
2 \left[1 - \exp \left(- L \frac{\Gamma_N}{\beta_N \gamma_N} \right) \right]
\nonumber\\  &&
+ 2 {\Bigg [}
{\bigg [} - \frac{(1-y^2)}{(1 + y^2)^2} \cos \left( 2\pi \frac{L}{L_{\rm osc}} \mp \Delta \psi^{\rm (X)} \right)
  + 2  \frac{y}{(1 + y^2)^2} \sin \left( 2\pi \frac{L}{L_{\rm osc}} \mp \Delta \psi^{\rm (X)} \right) {\bigg ]}
 \exp \left(- L \frac{\Gamma_N}{\beta_N \gamma_N} \right)
 \nonumber\\  &&
 + \left[
 \frac{(1-y^2)}{(1 + y^2)^2}  \cos(\Delta \psi^{\rm (X)})
 \pm 2 \frac{y}{(1 + y^2)^2} \sin (\Delta \psi^{\rm (X)}) \right] {\Bigg ]} {\Bigg \}}.
\label{Geffbexpr}
\eea \ees
We recall that X=LNC or X=LNV.
As mentioned, above we assumed that the speed $\beta_N$ (of $N_j$ in the lab frame $\Sigma$) is fixed. However, in practice this speed is not fixed. What is fixed is the speed $\beta'_N$ of $N_j$ in the $\tau$-rest frame $\Sigma'$.
\bes
\label{kinSigp}
\bea
E_N^{'} & = & \frac{(M_{\tau}^2 + M_N^2 - M_{\pi}^2)}{2 M_{\tau}}, \; |{\vec p'}_N|= \frac{1}{2} M_{\tau} \lambda^{1/2}\left( 1, \frac{M_N^2}{M_{\tau}^2}, \frac{M_{\pi}^2}{M_{\tau}^2} \right) ,
\label{ENppNp} \\
\beta_N^{'} \gamma^{'}_N &=& \sqrt{(E_N^{'}/M_N)^2 - 1} =
\frac{1}{2 M_N M_{\tau}} \left[ \left( (M_{\tau}+M_N)^2 - M_{\pi}^2 \right)
  \left( (M_{\tau}-M_N)^2 - M_{\pi}^2 \right) \right]^{1/2}.
\label{bNgNp} \eea \ees
On the other hand, it is realistic to assume that the velocity (the speed and the direction) ${\vec \beta}_{\tau}$ of the produced tau leptons in the lab frame $\Sigma$ is approximately fixed. We name this direction as the $z$-axis direction in the $N_j$-rest frame $\Sigma^{'}$, i.e., ${\hat \beta}_{\tau} \equiv {\hat z}^{'}$. In the transition  from the $\tau$-rest frame ($\Sigma^{'}$) to the lab frame ($\Sigma$), the resulting quantities $E_N$ and $\beta_N \gamma_N$ there will depend on the angle $\theta_N$ between the ${\vec p'}_N$ and ${\hat z}^{'}$ ($={\hat \beta}_{\tau}$) (cf.~Fig.~\ref{FigSpS})
\bes
\label{kinSig}
\bea
E_N & = & \gamma_{\tau} (E^{'}_N + \cos \theta_N \beta_{\tau} |{\vec p'}_N|),
\label{EN} \\
\beta_N \gamma_N &=&
\sqrt{ \gamma_{\tau}^2 \left( \frac{E^{'}_N + \cos \theta_N \beta_{\tau} |{\vec p'}_N|}{M_N} \right)^2 - 1} = \beta_N \gamma_N(\theta_N).
\label{bNgN} \eea \ees
Here, $E^{'}_N$ and $|{\vec p'}_N|$ are the constants given in Eq.~(\ref{ENppNp}). Similar considerations were made, in somewhat different contexts, in Ref.~\cite{Cvetic:2021itw} and in Ref.~\cite{GCCSK2017} (in this reference the lab frame was denoted as $\Sigma^{''}$).
As the angle $\theta_N$ (between ${\hat p}_{\tau} \equiv {\hat z}^{'}$ and ${\hat p'}_N$) appears in $\beta_N \gamma_N$, according to Eq.~(\ref{bNgN}), and this $\theta_N$ varies (${\hat p'}_N$ varies), we can see on inspection of the expression (\ref{Geffbexpr}) that this expression has the $\theta_N$-direction dependence only in the Lorentz factor $1/(\gamma_N \beta_N)$, and this factor appears in the exponent $\exp(- L \Gamma_N/(\gamma_N \beta_N))$ (in two places) and in $(2 \pi/L_{\rm osc})$, cf. Eq.~(\ref{Losc}). Using $d \Omega_{{\hat p'}_N} = 2 \pi d(\cos \theta_N)$), we can then write the final expression for the considered effective decay width
\bes
\label{Geff}
\bea
\lefteqn{
\Gamma_{\rm eff,\mp}^{\rm (X)} \equiv
\Gamma_{\rm eff}(\tau^{\mp} \to \pi^{\mp} \mu \pi; L)^{\rm (X)} =
\int_{0}^{L} d L' \int d \Omega_{{\hat p'}_N} \frac{d^2 \Gamma_{\rm eff}(\tau^{\mp} \to \pi^{\mp} \mu \pi; L')}{d L' d \Omega_{{\hat p'}_N}}
}
  \label{Geffdef} \\
  & = &
        {\overline \Gamma}(\tau \to \pi \mu \pi)  \; |U_{\tau N}|^2 |U_{\mu N}|^2 \times
  \int_{\cos \theta_N=-1}^{+1} d \cos \theta_N
 \nonumber\\ && \times
{\Bigg \{}
\left[1 -   \exp \left(- L \frac{\Gamma_N}{\beta_N \gamma_N(\theta_N)} \right) \right]
\nonumber\\  &&
+  {\Bigg [}
  {\bigg [}
- \frac{(1-y^2)}{(1 + y^2)^2}
\cos \left( 2\pi \frac{L}{L_{\rm osc}(\theta_N)} \right) \cos (\Delta \psi^{\rm (X)})
\mp \frac{(1-y^2)}{(1 + y^2)^2} \sin \left( 2\pi \frac{L}{L_{\rm osc}(\theta_N)} \right) \sin (\Delta \psi^{\rm (X)})
 \nonumber\\  &&
 + 2 \frac{y}{(1 + y^2)^2}  \sin \left( 2\pi \frac{L}{L_{\rm osc}(\theta_N)} \right) \cos (\Delta \psi^{\rm (X)})
 \mp 2  \frac{y}{(1 + y^2)^2} \cos \left( 2\pi \frac{L}{L_{\rm osc}(\theta_N)} \right) \sin (\Delta \psi^{\rm (X)})
 {\bigg ]}
  \exp \left(- L \frac{\Gamma_N}{\beta_N \gamma_N(\theta_N)} \right)
 \nonumber\\  &&
 + \left[
 \frac{(1-y^2)}{(1 + y^2)^2}  \cos(\Delta \psi^{\rm (X)})
 \pm 2 \frac{y}{(1 + y^2)^2} \sin (\Delta \psi^{\rm (X)}) \right] {\Bigg ]} {\Bigg \}}.
\label{Geffexpr}
\eea \ees
Here it is understood that the Lorentz factor product $\beta_N \gamma_N (\theta_N)$ is the expression (\ref{kinSig}) [in conjunction with Eqs.~(\ref{kinSigp})], and the integration over $d \cos \theta_N$ is to be performed numerically. Further, we recall that X=LNC (in that case: $\tau^{\mp} \to \pi^{\mp} \mu^{\mp} \pi^{\pm}$) or X=LNV ($\tau^{\mp} \to \pi^{\mp} \mu^{\pm} \pi^{\mp}$). The distance $L$ can be taken as any distance between zero and an effective detector length $L_{\rm det}$ ($\sim 1$ m). We will simply identify $L = L_{\rm det}$.

From the above expression, we can directly obtain the CP-asymmetry width
\bes
\label{GCP}
\bea
\Delta \Gamma^{\rm (X)}_{\rm CP}(L) & \equiv & \Gamma_{\rm eff}(\tau^- \to \pi^- \mu \pi; L)^{\rm (X)} - \Gamma_{\rm eff}(\tau^+ \to \pi^+ \mu \pi; L)^{\rm (X)}
\label{GCPdef}
\\
& = &
 {\overline \Gamma}(\tau \to \pi \mu \pi)  \; |U_{\tau N}|^2 |U_{\mu N}|^2 2  \sin (\Delta \psi^{\rm (X)}) \times
  \int_{\cos \theta_N=-1}^{+1} d \cos \theta_N
 \nonumber\\ && \times
 {\Bigg \{}
 \left[
- \frac{(1-y^2)}{(1 + y^2)^2} \sin \left( 2\pi \frac{L}{L_{\rm osc}(\theta_N)} \right)
-2  \frac{y}{(1 + y^2)^2} \cos \left( 2\pi \frac{L}{L_{\rm osc}(\theta_N)}  \right)
\right]
 \exp \left(- L \frac{\Gamma_N}{\beta_N \gamma_N(\theta_N)} \right)
\nonumber\\ &&
 + 2 \frac{y}{(1 + y^2)^2}.
 {\Bigg \}}
 \label{GCPexp} \eea \ees
This width can be regarded as a measure of CP violation in the considered LNV or LNC decays.

\section{Numerical results}
\label{sec:res}

The Belle II experiment \cite{Belle-II} is expected to produce about $N_{\tau} \sim 10^{11}$ lepton pairs $\tau^+ \tau^-$. An advantage of this experiment is that it can measure time dependent production rate. At Belle II the main source of $\tau$ pairs is the process $e^+ e^- \to \tau^+ \tau^-$. For this process, the lab energy of $\tau$'s is $E_{\tau} =5.29$ GeV, thus the lab velocity is $\beta_{\tau}=0.9419$ and $\gamma_{\tau} = 2.9772$.

\begin{figure}[htb] 
\begin{minipage}[b]{.49\linewidth}
\includegraphics[width=85mm,height=50mm]{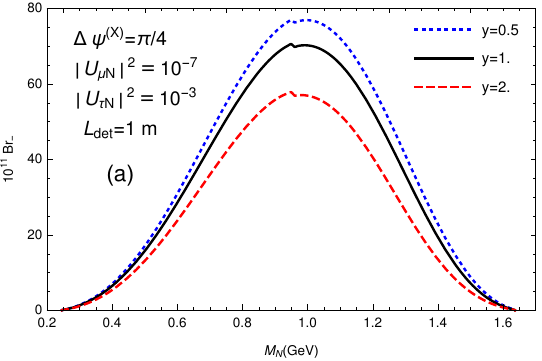}
\end{minipage}
\begin{minipage}[b]{.49\linewidth}
\includegraphics[width=75mm,height=47mm]{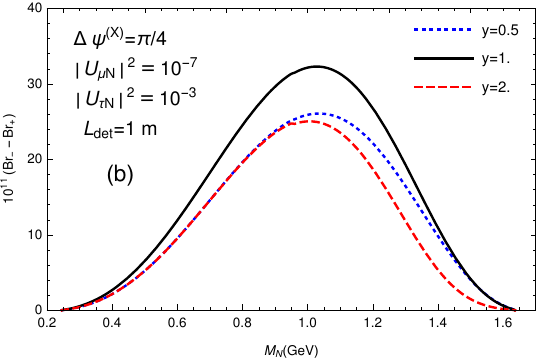}
\end{minipage} \vspace{12pt}
\caption{\footnotesize (a) The expected number of detected rare decays $\tau^- \to \pi^- N_j \to \pi^- \mu \pi$ [cf.~Eqs.~(\ref{Geff}) and (\ref{Nexp})] as a function of heavy neutrino mass $M_N$, for various values of $y= \Delta M_N/\Gamma_N$. The other parameters were fixed as indicated in the Figure (X=LNV or LNC). (b) The same, but now for the difference of the number of rare decays $\tau^- \to \pi^- \mu \pi$ and $\tau^+ \to \pi^+ \mu \pi$, cf.~Eqs.~(\ref{GCP}) and (\ref{Nexp}).}
\label{Figy}
\end{figure}

For the number of expected events of the considered decays, we have to evaluate the above rare decay widths $\Gamma_{\rm eff,\mp}^{\rm (X)}$, Eq.~(\ref{Geff}), divide it by the total $\tau$ decay width $\Gamma_{\tau}=2.267 \times 10^{-12}$ GeV to obtain the corresponding branching ratio ${\rm Br}_{\mp}$, and multiply this by the expected number $N_{\tau}=10^{11}$ of $\tau^-$ (or $\tau^+$) in order to get the expected number $N_{\rm exp.}^{\rm (X)}(\tau^{\mp})$ of such rare decays at Belle II (assuming no significant suppression from acceptance factors)
\be
N_{\rm exp.}(\tau^{\mp})^{\rm (X)} = {\rm Br}_{\mp}^{\rm (X)} \times 10^{11} =
  \frac{\Gamma_{\rm eff,\mp}^{\rm (X)}}{\Gamma_{\tau}}  \times 10^{11}.
  \label{Nexp} \ee
In Figs.~\ref{Figy}(a),(b) we present the quantities $N_{\exp}(\tau^-)^{\rm (X)}$ and $N_{\exp}(\tau^-)^{\rm (X)}-N_{\exp}(\tau^+)^{\rm (X)}$, respectively, as a function of heavy neutrino mass $M_N$, for various values of $y=\Delta M_M/\Gamma_N$ parameter, while keeping other parameters fixed as indicated in the Figure.
\begin{figure}[htb] 
\begin{minipage}[b]{.49\linewidth}
\includegraphics[width=85mm,height=50mm]{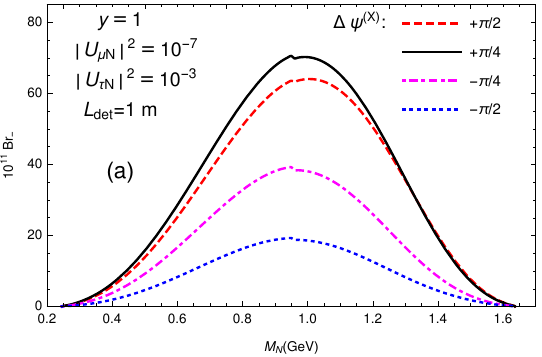}
\end{minipage}
\begin{minipage}[b]{.49\linewidth}
\includegraphics[width=75mm,height=47mm]{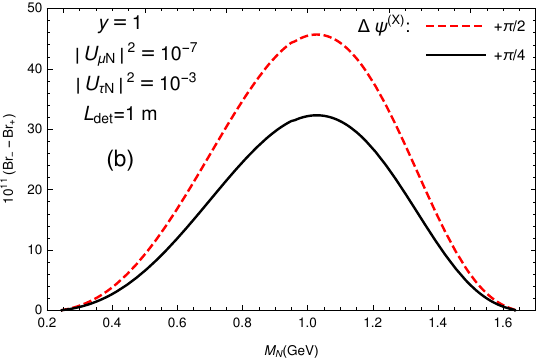}
\end{minipage} \vspace{12pt}
\caption{\footnotesize As Figs.~\ref{Figy}, but now for various values of the phase difference $\Delta \psi^{\rm (X)}$. In Fig.~(b) we took $\Delta \psi^{\rm (X)}$ only positive, because the depicted CP asymmetry is odd (changes sign) when  $\Delta \psi^{\rm (X)}$ changes sign, cf.~Eq.~(\ref{GCP}).}
\label{FigdPsi}
\end{figure}
\begin{figure}[htb] 
\begin{minipage}[b]{.49\linewidth}
\includegraphics[width=85mm,height=50mm]{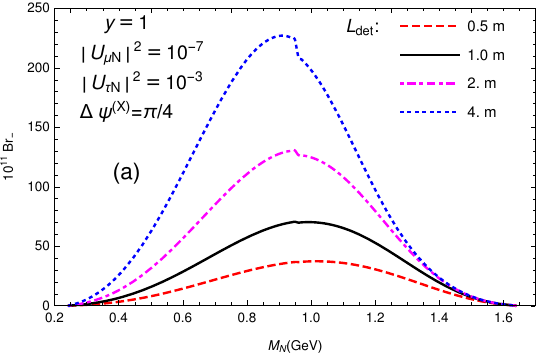}
\end{minipage}
\begin{minipage}[b]{.49\linewidth}
\includegraphics[width=75mm,height=47mm]{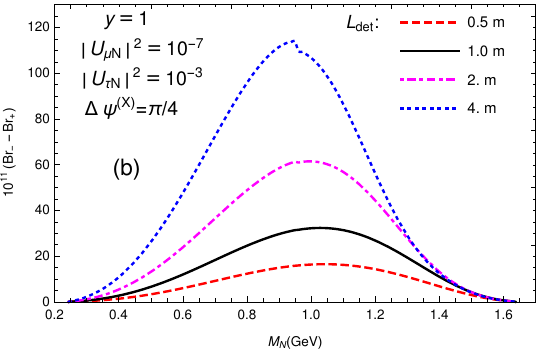}
\end{minipage} \vspace{12pt}
\caption{\footnotesize As Figs.~\ref{Figy}, but now for various values of the effective detector width $L_{\rm det}$.}
\label{FigLdet}
\end{figure}
\begin{figure}[htb] 
\begin{minipage}[b]{.49\linewidth}
\includegraphics[width=85mm,height=50mm]{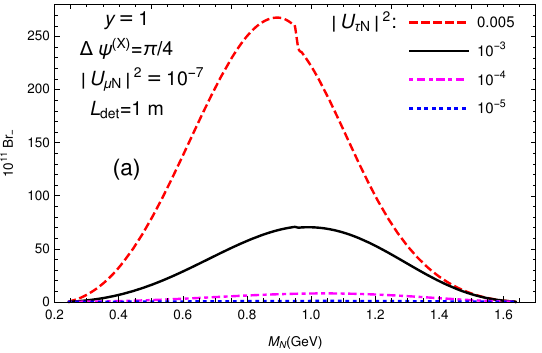}
\end{minipage}
\begin{minipage}[b]{.49\linewidth}
\includegraphics[width=75mm,height=47mm]{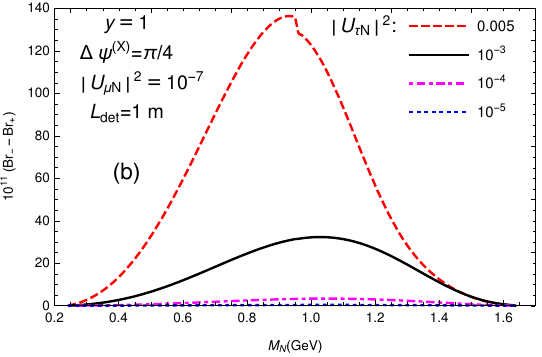}
\end{minipage} \vspace{12pt}
\caption{\footnotesize As Figs.~\ref{Figy}, but now for various values of the heavy-light mixing parameter $|U_{\tau N}|^2$.}
\label{FigUtaN2}
\end{figure}
As the central case, we take the values $y=1$, $\Delta \psi^{\rm (X)}=\pi/4$, $L_{\rm det}=1$ m, $|U_{\mu N}|^2=10^{-7}$ and $|U_{\tau N}|^2=10^{-3}$. The last two values are allowed by the present upper bounds on the heavy-light mixing parameters in the on-shell mass range $0.2 \ {\rm GeV} < M_N < 1.7$ GeV, cf.~Refs.~\cite{Atre,Cvetic:2018elt}.
The kink at $M_N=M_{\eta'}$ ($=0.9578$ GeV) in the presented curves appears due to the aforementioned kink in the calculated total decay width $\Gamma_N$ at that $M_N$ due to small duality violation effects, cf.~Figs.~\ref{FigGN} and discussion there.

In Figs.~\ref{FigdPsi} we present the analogous results, but now for various values of the phase differences $\Delta \psi^{\rm (X)}$. In Figs.~\ref{FigLdet} the effective length of the detector, $L_{\rm det}$, is varied. In Figs.~\ref{FigUtaN2} the heavy-light mixing parameter $|U_{\tau N}|^2$ is varied.
Further, it turns out that variation of the other heavy-light mixing parameter, $|U_{\mu N}|^2$, gives practically the results which are proportional to $|U_{\mu N}|^2$, the proportionality being valid by better than one percent precision.

In our evaluations, we used for $\Gamma_N$ the expressions for the Majorana neutrinos. However, it turns out that in the considered cases $\Gamma_N$ is practically equal for the Majorana and for the Dirac cases. Namely, $\Gamma_N$ depends almost entirely on the term ${\cal N}_{\tau N} |U_{\tau N}|^2$, cf.~Eqs.~(\ref{GammaNj})-(\ref{calKj}). This is so because of a set of two facts: (a) We took $|U_{e N}|^2=0$ (due to the $0 \nu 2 \beta$ restrictions), and we have $|U_{\mu N}|^2 \ll  |U_{\tau N}|^2$ in all cases of our choice of the numerical values of these heavy-light mixing parameters, due to the presently known bounds \cite{Atre,Cvetic:2018elt} (in the central case we took  $|U_{\mu N}|^2 = 10^{-7}$ and $|U_{\tau N}|^2 = 10^{-3}$); (b) In the considered mass range $0.2 \ {\rm GeV} < M_N < 1.7$ GeV, the dimensionless coefficient ${\cal N}_{\tau N}(M_N)$ has practically equal values for Majorana and Dirac (cf.~Figs.~\ref{FigGN}). The latter fact has its reason in the equality of the channels for the corresponding decays (in Majorana and Dirac case) that contribute to  ${\cal N}_{\tau N}(M_N)$. Namely, those channels are all of the form $N \to \nu_{\tau} l^{'-} l^{'+}$,  $N \to \nu_{\tau} \nu' {\bar \nu'}$, $N \to \nu_{\tau} V^0$, $N \to \nu_{\tau} P^0$, $N \to \nu_{\tau} q {\bar q}$, i.e., the channels that are unchanged under the charge-conjugation if $N$ is Majorana, and thus in Majorana case they do not get their value doubled in comparison with Dirac case. On the other hand, the relevant channels that change under the carge-conjugation are of the form $N \to \tau^{\mp} \ldots$, but these channels are kinematically not allowed in our considered cases (cf.~App.~A.3 of Ref.~\cite{symm}).

As a consequence, the numerical results for the branching ratios ${\rm Br}$ for the Dirac case practically do not differ from those of the Majorana case, for the same choice of the values of parameters $y$, $\Delta \psi^{\rm (LNC)}$, $|U_{\mu N}|^2$, $|U_{\tau N}|^2$, and $L_{\rm det}$. We do have to recall, though, that in the Dirac case only X=LNC option is realised, while in the Majorana case both X=LNC and X=LNV options are in general realised.

\section{Conclusions}
\label{sec:concl}

We derived the expressions for the effective widths of rare decays of $\tau$ leptons,
$\Gamma^{\rm (X)}_{{\rm eff}, \mp}  \equiv \Gamma_{\rm eff}(\tau^{\mp} \to \pi^{\mp} N_j \to \pi^{\mp} \mu \pi)^{\rm (X)}$, in the scenario where we have two on-shell almost mass degenerate neutrinos $N_j$ ($j=1,2$), and the decays are lepton number conserving (X=LNC: $\mu \mapsto \mu^{\mp}$) or lepton number violating (X=LNV: $\mu \mapsto \mu^{\pm}$). In the described almost degenerate scenario ($\Delta M_N \sim \Gamma_N$), we can have important $N_1$-$N_2$ oscillation effects, which also affect the CP-violating decay width $\Gamma^{\rm (X)}_{{\rm eff},-} - \Gamma^{\rm (X)}_{{\rm eff},+}$ where the latter is proportional to $\sin( \Delta \Psi^{\rm (X)})$. Here, $\Delta \Psi^{\rm (X)}$ is a specific phase difference of the heavy-light mixing parameters $U_{\mu N_j}$ and $U_{\tau N_j}$, cf.~Eqs.~(\ref{psi}) and (\ref{Delpsi}). In our expressions, we took into account the $N_1$-$N_2$ overlap and oscillation effects. Further, we accounted for the effects of the finite effective length $L_{\rm det}$ of the detector on the decay probability of the on-shell neutrinos within the detector. Subsequently, these decay widths and their branching ratios were numerically evaluated, for various values of the parameters of $y \equiv \Delta M_N/\Gamma_N$,  $\Delta \Psi^{\rm (X)}$, $L_{\rm det}$, $|U_{\mu N_j}|^2$, $|U_{\tau N_j}|^2$. For the Belle II experiment, where $\sim 10^{11}$ pairs of $\tau$-leptons are expected to be produced, we obtained that the expected number of such rare decays, $10^{11} {\rm Br}_{\mp}^{\rm (X)}$, can reach the order of $\sim 10^2$, and even the CP-violating difference of the number of events, $10^{11} ({\rm Br}_{-}^{\rm (X)}-{\rm Br}_{+}^{\rm (X)})$, can reach $\sim 10^2$. This implies that, if such a system of two almost degenerate heavy neutrinos with masses $0.3 \ {\rm GeV} < M_N < 1.5$ GeV exists, Belle II experiments could give us an indication of the existence of such a system by detecting a sufficient number of such decays. We note that the squared invariant mass of produced $\mu \pi$ is in such decays fixed, equal to $M_N^2$, thus leading to an identification of the mediating heavy neutral leptons $N_j$. Another possible identification would be via the localisation of the two vertices in the decay process. Further, in such cases, the indication of CP violation in the sector of such heavy neutrinos could be detected [$({\rm Br}_{-}^{\rm (X)}-{\rm Br}_{+}^{\rm (X)}) \propto \sin(\Delta \psi^{(X)})$]. If such decays are not detected, upper bound limits on the product $|U_{\mu N}|^2 |U_{\tau N}|^2$ could be refined.

Furthermore, if such rare decays are detected in both LNC and LNV modes, this would be a clear indication of the Majorana character of the neutrinos. If only LNC decays are detected, then this would indicate strongly that the neutrinos are Dirac.

The mathematica programs used to generate the presented curves in Figs.~\ref{FigGN}, \ref{Figy}-\ref{FigUtaN2}, are available \cite{www} on the web page \mbox{http://www.gcvetic.usm.cl/}.

\begin{acknowledgments}
\noindent
This work of G.C.was supported in part by FONDECYT (Chile) Grant No.~1220095,  The work of C.S.K. was supported by NRF of Korea (NRF-2021R1A4A2001897
and NRF-2022R1I1A1A01055643).
\end{acknowledgments}


\appendix

\section{$T_2^{(\rm X, 0)}$ and $T_2^{(\rm X, 1)}$ expressions, for X=LNC and X=LNV}
\label{app:T2X01}

The expressions $T_2^{(\rm LNC, 0)}$ and  $T_2^{(\rm LNC, 1)}$ apppearing in Eqs.~(\ref{T2LC}), are
\bes
\label{T2LC01}
\bea
T_2^{(\rm LNC, 0)} & = & -4  {\Bigg \{} M_N^6 M_{\pi}^2 - M_N^4 M_{\pi}^4 +
M_{\mu}^4 \left[ M_{\pi}^2 M_{\tau}^2 - (-M_N^2 + M_{\tau}^2)^2 \right]
\nonumber \\ &&
+
   M_{\mu}^2 \left[  -M_{\pi}^4 M_{\tau}^2 + (M_N^3 - M_N M_{\tau}^2)^2 +
      M_{\pi}^2 (-3 M_N^2 M_{\tau}^2 + M_{\tau}^4) \right]  {\Bigg \}}
\label{T2LC0} \\
T_2^{(\rm LNC, 1)} & = & - 8 M_N^2 (M_N^2 - M_{\mu}^2) (M_N^2 - M_{\tau}^2)
\label{T2LC1} \eea \ees
The expressions $T_2^{(\rm LNV, 0)}$ and  $T_2^{(\rm LNV, 1)}$ apppearing in Eqs.~(\ref{T2LV}), are
\bes
\label{T2LV01}
\bea
T_2^{(\rm LNV, 0)} & = & 4 M_N^2
 {\Bigg \{} -M_{\mu}^4 M_{\pi}^2 + M_{\pi}^4 M_{\tau}^2 + (M_N^3 - M_N M_{\tau}^2)^2 -
 M_{\pi}^2 (M_N^4 - M_N^2 M_{\tau}^2 + M_{\tau}^4)
\nonumber \\ &&
 +
   M_{\mu}^2
   \left[ M_{\pi}^4 - (-M_N^2 + M_{\tau}^2)^2 + M_{\pi}^2 (M_N^2 + M_{\tau}^2) \right]
   {\Bigg \}}
   \label{T2LV0} \\
T_2^{(\rm LNV, 1)} & = & 8 M_N^2 ( M_N^2 -M_{\mu}^2) (M_N^2 - M_{\tau}^2)
\label{T2LV1} \eea \ees

\section{Explicit integrations over final phase space}
\label{app:d3}

Using the factorisation Eq.~(\ref{d3}) of the final phase space integration differential, and using the final phase space integration formulas of Ref.~\cite{CollPhys}, we obtain
\bea
\int d_3 \times 1  &=& \int  d_2(\tau(p_{\tau}) \to \pi(p_1) N_j(p_N)) \; d p_N^2 \; d_2(N_j(p_N) \to \mu(p_2) \pi(p_{\pi})
\nonumber\\
& = & \frac{1}{8^2} \lambda^{1/2} \left( 1, \frac{M_N^2}{M_{\tau}^2}, \frac{M_{\pi}^2}{M_{\tau}^2} \right) \lambda^{1/2} \left( 1, \frac{M_{\mu}^2}{M_N^2}, \frac{M_{\pi}^2}{M_N^2} \right) \int d p_N^2 \int d \Omega_{{\hat p}_N^{'}} \int d \Omega_{{\hat p}_2^{''}}
\nonumber\\
& = & \frac{\pi^2}{4} \lambda^{1/2} \left( 1, \frac{M_N^2}{M_{\tau}^2}, \frac{M_{\pi}^2}{M_{\tau}^2} \right) \lambda^{1/2} \left( 1, \frac{M_{\mu}^2}{M_N^2}, \frac{M_{\pi}^2}{M_N^2} \right) \int d p_N^2 ,
\label{int1d3} \eea
where ${\hat p}_N^{'}$ is the unitary direction of $p_N$ in the CMS of $\tau$ ($\Sigma^{'}$ frame), and ${\hat p}_2^{''}$ is the unitary direction of $p_2$ (muon) in the CMS of $N_j$ ($\Sigma^{''}$ frame). The two integrations over these two directions give us factor $(4 \pi)^2$. The remaining integration over $p_N^2$ is over the kinematically allowed region $(M_{\mu}+M_{\pi})^2 \leq p_N^2 \leq (M_{\tau}-M_{\pi})^2$, but in practice it gives unity times the integrand in which we replace $p_N^2 \mapsto M_N^2$ because of the $\delta(p_N^2-M_N^2)$ factor appearing in the integrand, cf.~Eqs.(\ref{PP})-(\ref{T2}). Furthermore, $\lambda^{1/2}$ in Eq.~(\ref{int1d3}) is square root of the function
\be
\lambda(x,y,z) = x^2 + y^2 + z^2 - 2 x y - 2 y z - 2 z x.
\label{lamb} \ee
Similarly, the integration with the integrand $(p_1 \cdot p_2)$ gives
\bea
\int d_3 \times (p_1 \cdot p_2)   &=& \int d_2(\tau(p_{\tau}) \to \pi(p_1) N_j(p_N)) \; d_2(N_j(p_N) \to \mu(p_2) \pi(p_{\pi})) \; (p_1)_{\alpha} (p_2)^{\alpha} \; d p_N^2
\nonumber\\
& = & \int d p_N^2 \int d_2(\tau(p_{\tau}) \to \pi(p_1) N_j(p_N))  \; (p_1)_{\alpha}
\int d_2(N_j(p_N) \to \mu(p_2) \pi(p_{\pi}))  \; (p_2)^{\alpha}
\nonumber\\
& = & \int d p_N^2  \int d_2(\tau(p_{\tau}) \to \pi(p_1) N_j(p_N)) \; (p_1)_{\alpha} (p_N)^{\alpha} \frac{\pi}{4} \lambda^{1/2} \left( 1, \frac{M_{\mu}^2}{p_N^2}, \frac{M_{\pi}^2}{p_N^2} \right) \left(1 + \frac{(M_{\mu}^2-M_{\pi}^2)}{p_N^2} \right)
\nonumber\\
& = & \frac{\pi^2}{24} \int d p_N^2 \;
\lambda^{1/2} \left( 1, \frac{p_N^2}{M_{\tau}^2}, \frac{M_{\pi}^2}{M_{\tau}^2} \right)\lambda^{1/2} \left( 1, \frac{M_{\mu}^2}{p_N^2}, \frac{M_{\pi}^2}{p_N^2} \right)
\left(1 + \frac{(M_{\mu}^2-M_{\pi}^2)}{p_N^2} \right) \times
\nonumber\\ &&
\left\{ \lambda \left( 1, \frac{p_N^2}{M_{\tau}^2}, \frac{M_{\pi}^2}{M_{\tau}^2} \right) + \frac{1}{2} \left[ 1 + \frac{(p_N^2+M_{\pi}^2)}{M_{\tau}^2} - 2 \frac{(p_N^2 - M_{\pi}^2)^2}{M_{\tau}^4} \right] \right\},
\label{intp1p2d3} \eea
where, as in Eq.~(\ref{int1d3}), the integration over $p_N^2$ again gives us unity times the integrand in which we replace $p_N^2 \mapsto M_N^2$, due to the relations Eqs.~(\ref{PP})-(\ref{T2}).

\end{document}